\documentclass[prd,superscriptaddress,amsfonts,amssymb,amsmath,showpacs,twocolumn,floatfix]{revtex4-2}
\usepackage{bm}
\usepackage{amsfonts}
\usepackage{latexsym}
\usepackage{graphicx}
\usepackage{amsmath}
\usepackage{palatino}
\usepackage{mathpazo}
\usepackage{textcomp}
\usepackage{comment}
\usepackage{float}
\usepackage{amsmath,amssymb}

\usepackage{booktabs}
\usepackage{dcolumn}
\usepackage{booktabs}
\usepackage{multirow}
\usepackage{amsmath}
\usepackage{xcolor}
\usepackage{orcidlink}
\usepackage[caption=false]{subfig}
\usepackage{commath}
\usepackage{tabularx}

\def\jnl@style{\it}
\def\aaref@jnl#1{{\jnl@style#1}}

\def\aaref@jnl#1{{\jnl@style#1}}

\def\aj{\aaref@jnl{AJ}}                   
\def\apj{\aaref@jnl{ApJ}}                 
\def\apjl{\aaref@jnl{ApJ}}                
\def\apjs{\aaref@jnl{ApJS}}               
\def\apss{\aaref@jnl{Ap\&SS}}             
\def\aap{\aaref@jnl{A\&A}}                
\def\aapr{\aaref@jnl{A\&A~Rev.}}          
\def\aaps{\aaref@jnl{A\&AS}}              
\def\mnras{\aaref@jnl{Mon.~Not.~Roy.~Astron.~Soc.}}             
\def\prd{\aaref@jnl{Phys.~Rev.~D}}        
\def\prc{\aaref@jnl{Phys.~Rev.~C}}  
\def\prl{\aaref@jnl{Phys.~Rev.~Lett.}}    
\def\qjras{\aaref@jnl{QJRAS}}             
\def\skytel{\aaref@jnl{S\&T}}             
\def\ssr{\aaref@jnl{Space~Sci.~Rev.}}     
\def\zap{\aaref@jnl{ZAp}}                 
\def\nat{\aaref@jnl{Nature}}              
\def\aplett{\aaref@jnl{Astrophys.~Lett.}} 
\def\apspr{\aaref@jnl{Astrophys.~Space~Phys.~Res.}} 
\def\physrep{\aaref@jnl{Phys.~Rep.}}      
\def\physscr{\aaref@jnl{Phys.~Scr}}       
\def\commat{\aaref@jnl{Comm.~Math.~Phys.}}              
\def\science{\aaref@jnl{Science}}               
\def\cqg{\aaref@jnl{Classical Quant.~Grav.}}            
\def\jpcs{\aaref@jnl{JPCS}}                                     
\def\ijmpd{\aaref@jnl{Int.~J.~Mod.~Phys.~D}}                    
\def\grg{\aaref@jnl{Gen.~Relat.~Gravit.}}               
\def\rpp{\aaref@jnl{Rep.~Prog.~Phys.}}          
\def\npa{\aaref@jnl{Nucl.~Phys.~A}}        
\def\lrr{\aaref@jnl{Living Rev.~Rel.}}                   
\def\jcap{\aaref@jnl{J.~Cosmology Astropart.~Phys.}}    
\def\rmp{\aaref@jnl{Rev.~Mod.~Phys.}}   
\def\epjc{\aaref@jnl{Eur.~Phys.~J.~C}}

\allowdisplaybreaks[1]
\renewcommand{\arraystretch}{1.1}
\usepackage{hyperref}
\hypersetup{colorlinks,citecolor=blue}
\begin{document}

\title{Probing dark energy dynamics through a new CPL-type $Om(z)$ parametrization in FLRW Universe}

\author{D. Revanth Kumar\orcidlink{0009-0002-6599-3608}}
\email{2406c9m003@sru.edu.in}
\affiliation{SR University, Warangal-506371, Telangana, India.}

\author{Santosh Kumar Yadav\orcidlink{0009-0009-2581-387X}}
\email[Corresponding author: ]{sky91bbaulko@gmail.com}
\affiliation{SR University, Warangal-506371,
Telangana, India.}

\author{S. A. Kadam\orcidlink{0000-0002-2799-7870}}
\email{siddheshwar.kadam@dypiu.ac.in;
\\k.siddheshwar47@gmail.com}
\affiliation{Centre for Interdisciplinary Studies and Research, D Y Patil International University, Akurdi, Pune-411044, Maharashtra, India.}


\begin{abstract}
In this work, we investigate a CPL-type $Om(z)$ parametrization given by $Om(z)=l+m\left(\frac{z}{1+z}\right)$ to study the evolution of dark energy and possible deviations from the standard $\Lambda$CDM cosmology. The model reduces to the flat $\Lambda$CDM scenario for $m=0$, while $m\neq0$ indicates dynamical evolution. The free parameters are constrained using Pantheon+SH0ES (PPS), Cosmic Chronometer (CC), and DESI BAO DR2 observations through Markov Chain Monte Carlo analysis.
The obtained constraints yield $72.48\lesssim H_0\lesssim72.97~\mathrm{km\,s^{-1}\,Mpc^{-1}}$, while the evolution parameter remains negative for all dataset combinations, leading to decreasing $Om(z)$ evolution and favoring quintessence-like behaviour. The combined PPS+CC+DR2 analysis shifts the evolution closer to the $\Lambda$CDM limit. The statistical analysis based on $\chi^2$, AIC, and BIC indicates that the proposed model remains statistically competitive with $\Lambda$CDM.
The deceleration parameter and total equation of state are further analyzed to examine the cosmic evolution. The transition from deceleration to acceleration occurs within $z_{tr}\approx0.66-1.01$, while the present values vary within $-0.459\lesssim q_0\lesssim-0.325$ and $-0.639\lesssim\omega_0\lesssim-0.550$ depending on the datasets. Overall, the proposed parametrization provides a flexible framework for studying late-time cosmic evolution.

\textbf{Keywords:} $Om(z)$ parametrization, observational constraints, Hubble constant, dark energy.
\end{abstract}

\maketitle

\section{Introduction}\label{sec1}

The accelerated expansion of the Universe remains one of the most intriguing discoveries in contemporary cosmology. Evidence for this phenomenon first emerged from observations of Type Ia supernovae (SNe Ia)~\cite{SupernovaSearchTeam:1998fmf,SupernovaCosmologyProject:1998vns} and was subsequently reinforced by several independent cosmological probes, including measurements of the cosmic microwave background (CMB) radiation~\cite{aghanim2020planck,WMAP:2003ivt}, baryon acoustic oscillations (BAO)~\cite{SDSS:2005xqv,SDSS:2009ocz}, and large-scale structure observations. The origin of this late-time acceleration remains unclear and is generally attributed to an unknown component with repulsive gravitational effects, commonly referred to as dark energy (DE)~\cite{Peebles:2002gy,Padmanabhan:2002ji,Copeland:2006wr}. Within the standard cosmological framework, the simplest realization of DE is provided by the cosmological constant $\Lambda$, leading to the $\Lambda$ cold dark matter ($\Lambda$CDM) model, where the equation of state (EoS) parameter remains fixed at $\omega=-1$.

Despite its remarkable agreement with a broad class of observations, the $\Lambda$CDM model is not free from conceptual and observational difficulties. One of the major theoretical challenges is associated with the cosmological constant problem, arising from the huge mismatch between the observed vacuum energy density and the value predicted by quantum field theory~\cite{Weinberg:1988cp,Carroll:1991mt}. Another unresolved issue is the coincidence problem~\cite{Zlatev:1998tr}, which questions why the energy densities of matter and DE become comparable only during the present cosmological epoch despite their distinct evolutionary behaviors. In addition to these theoretical issues, recent cosmological observations have revealed persistent tensions in parameter estimation, most notably in the determination of the Hubble constant $H_0$. Measurements based on local observations generally prefer larger values of $H_0$, whereas CMB-inferred estimates within the $\Lambda$CDM framework favor smaller values~\cite{aghanim2020planck,riess2022comprehensive,Verde:2019ivm,Knox:2019rjx,Kamionkowski:2022pkx,Riess:2024vfa}. This discrepancy, usually referred to as the \textit{Hubble tension}, together with several recently reported cosmological anomalies~\cite{Rogers:2023upm,Ruchika:2024ymt,Lopez-Corredoira:2024pgl,Pourojaghi:2024bxa,Green:2024xbb,Khalife:2023qbu,Hazra:2024nav}, has motivated extensive investigations of alternatives beyond the standard cosmological paradigm.

To explain the accelerated expansion of the Universe, a large number of theoretical scenarios have been proposed. Broadly, these approaches can be divided into two categories. The first category modifies the gravitational sector itself, leading to different extensions of General Relativity~\cite{Copeland:2006wr,Clifton:2011jh,Koyama:2015vza,nojiri2017modified,koussour2023constraining,duchaniya2024cosmological, Kadam:2026pjm}. The second category retains the standard gravitational framework but introduces dynamical DE components with evolving properties~\cite{DeFelice:2010aj, Scherer:2025esj, Giare:2024gpk, Yadav:2019jio, nagpal2025late, Escamilla:2024fzq,Pacif:2020hai,
abdalla2022cosmology, RevanthKumar:2026ayb}. Several DE candidates have been explored in this direction, including quintessence fields~\cite{Capozziello:2002rd,Zimdahl:2001ar,Amendola:1999er}, phantom scenarios~\cite{Nojiri:2005sx,Caldwell:2003vq}, k-essence models~\cite{Armendariz-Picon:2000nqq}, Chaplygin gas cosmologies~\cite{Kamenshchik:2001cp}, tachyon fields~\cite{Sen:2002in}, and holographic DE models~\cite{Li:2004rb, Anagnostopoulos:2020ctz}. Since many of these models predict an evolving EoS parameter, determining whether DE evolves dynamically remains a central objective in modern cosmology.

Among the different approaches developed to investigate the evolution of DE, the $Om(z)$ diagnostic has emerged as an efficient and model-independent tool for distinguishing the cosmological constant from dynamical DE scenarios. The diagnostic was originally proposed by Sahni \textit{et al.}~\cite{Sahni:2008xx} as a null test of the $\Lambda$CDM model and has since been widely employed to examine possible deviations from the standard cosmological framework.

One of the major advantages of the $Om(z)$ diagnostic is its capability to characterize DE evolution directly through the behavior of the Hubble parameter without relying on a specific DE model. A constant $Om(z)$ throughout cosmic evolution remains consistent with the cosmological constant scenario, whereas a varying $Om(z)$ indicates departures from $\Lambda$CDM and points toward dynamical  nature of DE. Furthermore, the slope of the diagnostic provides additional information regarding the nature of DE. In particular, a positive slope generally indicates phantom-like evolution $(\omega_{DE}<-1)$, while a negative slope is associated with quintessence behavior $(\omega_{DE}>-1)$.

Motivated by these properties, several studies have employed the $Om(z)$ diagnostic using different observational probes and reconstruction techniques. Sahni \textit{et al.}~\cite{Sahni:2014ooa} utilized an improved version of the $Om(z)$ diagnostic together with BAO observations and reported evidence supporting evolving DE beyond the standard $\Lambda$CDM scenario. Ding \textit{et al.}~\cite{Ding:2015vpa} investigated the evolution of DE by combining gravitational lensing observations, SNe Ia, and BAO measurements within the $Om(z)$ framework. Zheng \textit{et al.}~\cite{Zheng:2016jlq} analyzed both $Om(z)$ and $Omh^2(z_1,z_2)$ diagnostics using observational Hubble data to examine possible deviations from the cosmological constant scenario. Seikel \textit{et al.}~\cite{Seikel:2012cs} further examined the consistency of the $\Lambda$CDM model through Gaussian-process reconstructions of $Om(z)$ using observational datasets.

In recent years, several phenomenological parametrizations of the $Om(z)$ diagnostic have been introduced to investigate the dynamical evolution of DE and possible departures from $\Lambda$CDM. One commonly adopted form is the power-law parametrization, $Om(z)=\alpha(1+z)^n$, proposed by Myrzakulov \textit{et al.}~\cite{Myrzakulov:2023rxp}, where the parameters $\alpha$ and $n$ characterize possible deviations from the standard cosmological scenario. Later, Myrzakulov \textit{et al.}~\cite{Myrzakulov:2025bfv} explored logarithmic forms of the type $Om(z)=\alpha \ln(1+z)+\beta$, to describe gradual low-redshift evolution and smooth departures from the $\Lambda$CDM model. Qi \textit{et al.}~\cite{Qi:2018pej} investigated a parametrization of the form $Om(z)=\alpha \left(\frac{z}{1+z}\right)^n$, which allows smooth interpolation between different cosmological epochs. More recently, Yadav \textit{et al.}~\cite{Yadav:2026jaw} proposed transition parametrizations of the form $Om(z)=\frac{z^l}{(1+z)^m}$, to investigate the cosmological evolution of DE across different redshift domains. Furthermore, Bag \textit{et al.}~\cite{Bag:2026eju} introduced the $w_0$-probe, a novel diagnostic constructed from the $Om(z)$ function that enables a model-independent determination of the present-day DE EoS parameter directly from the expansion history, further demonstrating the growing importance of $Om(z)$-based approaches in probing DE dynamics.

Following these recent investigations, we have introduced a new Chevallier-Polarski-Linder (CPL)~\cite{chevallier2001accelerating, linder2003exploring} inspired parametrization of the $Om(z)$ diagnostic given by $Om(z)=l+m\left(\frac{z}{1+z}\right)$, where $l$ and $m$ are free model parameters constrained using observational datasets. The proposed form is motivated by the CPL framework, where the factor $z/(1+z)$ enables a smooth interpolation between different cosmological epochs while maintaining mathematical simplicity. Unlike several existing parametrizations, the present form provides a flexible framework to investigate different evolutionary behaviors of the $Om(z)$ diagnostic and possible deviations from the standard $\Lambda$CDM scenario. To assess the viability of the proposed model, we perform observational constraints using Cosmic Chronometer (CC) measurements, Pantheon+ Type Ia supernova data, and DESI DR2 BAO observations. The evolution of the cosmological parameters is investigated, and the implications of the reconstructed dynamics for the late-time Universe are discussed.

The paper is organized as follows. In Section~\ref{sec2}, we introduce the cosmological framework and derive the model equations corresponding to the proposed $Om(z)$ parametrization. The observational datasets and statistical methodology employed in the analysis are presented in Section~\ref{sec3}. The observational constraints and cosmological implications are discussed in Section~\ref{sec4}. Finally, the main conclusions of the present work are summarized in Section~\ref{sec5}.

\section{Cosmological framework and model formulation}
\label{sec2}

In this section, we present the cosmological framework and derive the governing equations corresponding to the proposed $Om(z)$ parametrization. We assume that the large-scale evolution of the Universe obeys the cosmological principle, according to which the Universe is homogeneous and isotropic on sufficiently large scales. Under this assumption, the spacetime geometry is described by the Friedmann--Lema\^{\i}tre--Robertson--Walker (FLRW) metric,
\begin{equation}
ds^2=-dt^2+a^2(t)
\left[
\frac{dr^2}{1-kr^2}
+r^2(d\theta^2+\sin^2\theta d\phi^2)
\right],
\label{metric}
\end{equation}
where $a(t)$ denotes the scale factor describing the expansion history of the Universe and $k$ represents the spatial curvature parameter. For simplicity, we restrict our analysis to a spatially flat FLRW Universe and adopt natural units $(8\pi G=c=1)$. The scale factor is normalized as $a_0=1$ at the present.

The matter content of the Universe is modeled as a perfect fluid with energy--momentum tensor
\begin{equation}
T_{\mu\nu}
=
(\rho+p)u_\mu u_\nu
-pg_{\mu\nu},
\label{emt}
\end{equation}
where $\rho$ and $p$ denote the total energy density and pressure of the cosmic fluid, respectively, and $u_\mu$ is the four-velocity.

The dynamics of the Universe are governed by the Einstein field equations,
\begin{equation}
R_{\mu\nu}
-\frac{1}{2}g_{\mu\nu}R = T_{\mu\nu},
\label{efe}
\end{equation}
where $R_{\mu\nu}$ and $R$ represent the Ricci tensor and Ricci scalar, respectively.

Using Eqs.~(\ref{metric})--(\ref{efe}), the Einstein equations reduce to the Friedmann equations for a spatially flat FLRW Universe,
\begin{equation}
\begin{aligned}
3H^2 &= \rho,\\
2\dot H+3H^2 &= -p.
\end{aligned}
\label{fr}
\end{equation}
where $H=\frac{\dot a}{a}$
is the Hubble parameter and the overdot denotes differentiation with respect to cosmic time.

The first Friedmann equation connects the expansion rate of the Universe with the total energy density, while the second Friedmann equation describes the influence of pressure on cosmic acceleration.
To characterize the expansion history and investigate the dynamical behavior of DE, we consider the total EoS parameter, $\omega=\frac{p}{\rho}$.
Using Eqs.~(\ref{fr}), the EoS parameter can be expressed as
\begin{equation}
\omega=-\frac{2\dot H+3H^2}{3H^2}=-1-\frac{2\dot H}{3H^2}.
\label{eos}
\end{equation}

The $Om(z)$ diagnostic is defined as
\begin{equation}
Om(z)=\frac{E^2(z)-1}{(1+z)^3-1},
\label{om}
\end{equation}
where $E(z)=\frac{H(z)}{H_0}$ represents the normalized Hubble parameter and $H_0$ denotes the present value of the Hubble constant.
Rearranging Eq.~(\ref{om}), we obtain
\begin{equation}
E^2(z)=Om(z)\left[(1+z)^3-1\right]+1.
\label{Ez}
\end{equation}
For the standard flat $\Lambda$CDM cosmology, the $Om(z)$ diagnostic remains constant throughout the cosmic evolution and is given by $Om(z)=\Omega_{m0}$, where $\Omega_{m0}$ denotes the present matter density parameter. Therefore, any deviation from a constant $Om(z)$ behaviour indicates possible dynamical evolution beyond the standard $\Lambda$CDM framework. Substituting $Om(z)=\Omega_{m0}$ into Eq.~(\ref{Ez}), we obtain the standard flat $\Lambda$CDM Hubble parameter as
\begin{equation}
H(z)=H_0\sqrt{1+\Omega_{m0}\left[(1+z)^3-1\right]}.
\end{equation}

Motivated by recent developments in the literature, we propose a new CPL-inspired parametrization of the $Om(z)$ diagnostic as follows
\begin{equation}
Om(z)=l+m\left(\frac{z}{1+z}\right),
\label{model}
\end{equation}
where $l$ and $m$ are free parameters constrained using observational datasets.

The proposed form follows the CPL framework, where the factor $z/(1+z)$ naturally provides a smooth interpolation between low and high-redshift regimes while remaining finite throughout cosmic evolution. This property enables a continuous description of cosmological dynamics across different epochs using a mathematically simple parametrization.
An additional advantage of the model is its minimal parameter structure. The parameter $l$ determines the present value of the diagnostic, whereas $m$ governs its cosmological evolution. Depending on the sign and magnitude of $m$, the model can accommodate both increasing and decreasing behaviors of $Om(z)$, thereby allowing the investigation of phantom-like and quintessence-like evolution within a unified framework.

Furthermore, the parametrization exhibits well-behaved asymptotic limits. At low redshift $(z\rightarrow0)$, the model reduces to $Om(z)\rightarrow l$, whereas at high redshift $(z\rightarrow\infty)$ it approaches $Om(z)\rightarrow l+m$. In addition, for the special case $m=0$, the proposed parametrization reduces to a constant form, $Om(z)=l$. Since a constant $Om(z)$ corresponds to the flat $\Lambda$CDM scenario, the parameter $l$ acts as the present matter density parameter, i.e., $l=\Omega_{m0}$.

Therefore, the proposed form naturally connects different cosmological epochs and provides an efficient framework to investigate possible deviations from the standard $\Lambda$CDM scenario.

Substituting Eq.~(\ref{model}) into Eq.~(\ref{Ez}), the dimensionless Hubble parameter becomes
\begin{equation}
E^2(z)=1+\left[l+m\left(\frac{z}{1+z}\right)\right]\left[(1+z)^3-1\right].
\label{Efinal}
\end{equation}
Thus, the Hubble parameter is obtained as
\begin{equation}
H(z)=H_0\sqrt{1+\left[l+m\left(\frac{z}{1+z}\right)\right]\left[(1+z)^3-1\right]}.
\label{Hz_model}
\end{equation}
The cosmological evolution strongly depends on the parameters $l$ and $m$. Therefore, constraining these parameters through observational datasets enables us to investigate the dynamical behavior of DE and possible deviations from the standard $\Lambda$CDM cosmology.

\section{Observational datasets and statistical methodology}
\label{sec3}

To constrain the free parameters of the proposed $Om(z)$ parametrization, we employ recent late-time cosmological observations including CC measurements of the Hubble parameter, Type Ia supernova observations from the Pantheon+SH0ES (\texttt{PPS}) compilation, and BAO measurements from DESI Data Release 2 (DR2). In the present analysis, we consider the combinations \texttt{CC+PPS} and \texttt{CC+PPS+DR2}. The joint analysis of these complementary datasets enables tighter constraints on the cosmological parameters and provides consistency tests for the proposed scenario.

The parameter estimation is carried out using the Markov Chain Monte Carlo (MCMC) method implemented through the publicly available \texttt{emcee} Python package~\cite{foreman2013emcee}. The posterior distributions are analyzed using the \texttt{GetDist} package~\cite{Lewis:2019xzd}. We employ 150 walkers with $50000$ iterations to ensure convergence anld efficient sampling of the parameter space. For our model, uniform priors are adopted as:  $40<H_0<100,\quad -2<l<5,\quad -2<m<10$. For the $\Lambda$CDM model, the matter density parameter is assigned the prior $0<\Omega_{m0}<0.6$. The convergence of the chains is verified through the Gelman--Rubin criterion with $R-1<0.01$.

\subsection{CC dataset}

The CC approach provides direct and nearly model-independent measurements of the Hubble parameter by estimating the differential ages of passively evolving galaxies~\cite{Simon:2004tf,Stern:2009ep,Moresco:2012jh,Moresco:2015cya,moresco20166,Ratsimbazafy:2017vga}. The Hubble parameter is related to the redshift evolution through $H(z)=-\frac{1}{1+z}\frac{dz}{dt}$. In this work, we use 31 CC measurements covering the redshift interval $0<z<1.965$, summarized in Table~\ref{CC}. 

\begin{table}[ht]
\centering
\caption{31 CC measurements employed in the present analysis}
\label{CC}

\setlength{\tabcolsep}{5pt}
\renewcommand{\arraystretch}{1.3}

\begin{tabular}{ccc|ccc}
\hline\hline
$z$ & $H(z)\pm\sigma_H$ & Ref. &
$z$ & $H(z)\pm\sigma_H$ & Ref.\\
\hline

0.07  & $69.0\pm19.6$ & \cite{zhang2016test} &
0.4783 & $80.9\pm9.0$ & \cite{moresco20166} \\

0.09  & $69.0\pm12.0$ & \cite{Simon:2004tf} &
0.48  & $97.0\pm62.0$ & \cite{Stern:2009ep} \\

0.12  & $68.6\pm26.2$ & \cite{zhang2016test} &
0.593 & $104.0\pm13.0$ & \cite{Moresco:2012jh} \\

0.17  & $83.0\pm8.0$ & \cite{Simon:2004tf} &
0.68  & $92.0\pm8.0$ & \cite{Moresco:2012jh} \\

0.179 & $75.0\pm4.0$ & \cite{Moresco:2012jh} &
0.781 & $105.0\pm12.0$ & \cite{Moresco:2012jh} \\

0.199 & $75.0\pm5.0$ & \cite{Moresco:2012jh} &
0.875 & $125.0\pm17.0$ & \cite{Moresco:2012jh} \\

0.20  & $72.9\pm29.6$ & \cite{zhang2016test} &
0.88  & $90.0\pm40.0$ & \cite{Stern:2009ep} \\

0.27  & $77.0\pm14.0$ & \cite{Simon:2004tf} &
0.90  & $117.0\pm23.0$ & \cite{Simon:2004tf} \\

0.28  & $88.8\pm36.6$ & \cite{zhang2016test} &
1.037 & $154.0\pm20.0$ & \cite{Moresco:2012jh} \\

0.352 & $83.0\pm14.0$ & \cite{Moresco:2012jh} &
1.30  & $168.0\pm17.0$ & \cite{Simon:2004tf} \\

0.3802 & $83.0\pm13.5$ & \cite{moresco20166} &
1.363 & $160.0\pm33.6$ & \cite{Moresco:2015cya} \\

0.4004 & $77.0\pm10.2$ & \cite{moresco20166} &
1.43 & $177.0\pm18.0$ & \cite{Simon:2004tf} \\

0.4247 & $87.1\pm11.2$ & \cite{moresco20166} &
1.53 & $140.0\pm14.0$ & \cite{Simon:2004tf} \\

0.4497 & $92.8\pm12.9$ & \cite{moresco20166} &
1.75 & $202.0\pm40.0$ & \cite{Simon:2004tf} \\

0.47 & $89.0\pm50.0$ & \cite{Ratsimbazafy:2017vga} &
1.965 & $186.5\pm50.4$ & \cite{Moresco:2015cya} \\

0.40 & $95.0\pm17.0$ & \cite{Simon:2004tf} &
& & \\

\hline\hline
\end{tabular}
\end{table}
The corresponding likelihood function is constructed through
\begin{equation}
\chi^2_{\rm CC}=\sum_{i=1}^{N_{\rm CC}} \frac{\Delta H_i^{\,2}} {\sigma_{H,i}^{\,2}},
\end{equation}
where $\Delta H_i=H_{\rm th}(z_i,\Psi)-H_{\rm obs}(z_i)$, represents the difference between the theoretical and observed Hubble parameter values at redshift $z_i$. Here, $\Psi=(H_0,l,m)$ denotes the set of free parameters of the proposed model.

\subsection{Pantheon+SH0ES}

Type Ia supernovae are widely used as standard candles owing to their nearly uniform peak luminosities and provide strong constraints on the late-time expansion history of the Universe. In the present analysis, we employ the \texttt{PPS} compilation~\cite{brout2022pantheon+}, consisting of 1701 light curves corresponding to 1550 spectroscopically confirmed SNe Ia in the redshift interval $0.001<z<2.26$. The inclusion of SH0ES cepheid calibrations improves the absolute magnitude calibration and strengthens constraints on late-time cosmological parameters. The apparent magnitude relation is written as
\begin{equation}
m_B-M_B=5\log_{10} \left[\frac{D_L(z)}{\rm Mpc} \right]+25,
\end{equation}
where $M_B$ is the absolute magnitude and the luminosity distance for a flat Universe is given by
\begin{equation}
D_L(z)=\frac{(1+z)}{H_0}\int_0^z\frac{dz'}{E(z')}.
\end{equation}
The theoretical distance modulus becomes
\begin{equation}
\mu_{\rm th}=5\log_{10} [D_L(z)]+25.
\end{equation}
The likelihood function for the \texttt{PPS} sample is evaluated through
\begin{equation}
\chi^2_{\rm PPS}=\sum_{i,j}^{1701}\Delta\mu_i(C^{-1}_{\rm stat+syst})_{ij}\Delta\mu_j,
\end{equation}
where $\Delta\mu_i=\mu_i^{\rm th}-\mu_i^{\rm obs}$,
and $C_{\rm stat+syst}$ denotes the full covariance matrix including both statistical and systematic contributions.

\subsection{DESI BAO DR2 dataset}

BAO provide a standard ruler for probing the expansion history of the Universe through the sound horizon at the drag epoch. The corresponding acoustic scale is given by
\begin{equation}
r_d=\int_{z_d}^{\infty}\frac{c_s(z)}{H(z)}\,dz,
\end{equation}
where $c_s(z)$ denotes the sound speed in the baryon--photon plasma and $z_d$ represents the drag epoch redshift.

In this work, we employ the recent BAO measurements from the DESI DR2 compilation~\cite{karim2025desi}. The dataset consists of nine measurements spanning the redshift interval $0.295<z<2.33$, providing constraints through the quantities $D_M/r_d$, $D_H/r_d$, and $D_V/r_d$, where $D_M$, $D_H$, and $D_V$ denote the transverse comoving distance, Hubble horizon distance, and angle-averaged distance, respectively. The observational values used in the present analysis are listed in Table~\ref{tab:DESI}. Throughout the paper, this dataset is referred to simply as \texttt{DR2}.
\begin{table}[ht]
\centering
\caption{9 DESI BAO DR2 measurements used in the present analysis.}
\label{tab:DESI}
\setlength{\tabcolsep}{5pt}
\renewcommand{\arraystretch}{1.3}
\begin{tabular}{cccc}
\hline\hline
$z_{\rm eff}$ & $D_M/r_d$ & $D_H/r_d$ & $D_V/r_d$ \\
\hline

0.295 & -- & -- &
$7.942\pm0.075$ \\

0.510 &
$13.588\pm0.167$ &
$21.863\pm0.425$ &
$12.720\pm0.099$ \\

0.706 &
$17.351\pm0.177$ &
$19.455\pm0.330$ &
$16.050\pm0.110$ \\

0.934 &
$21.576\pm0.152$ &
$17.641\pm0.193$ &
$19.721\pm0.091$ \\

0.922 &
$21.648\pm0.178$ &
$17.577\pm0.213$ &
$19.656\pm0.105$ \\

0.955 &
$21.707\pm0.335$ &
$17.803\pm0.297$ &
$20.008\pm0.183$ \\

1.321 &
$27.601\pm0.318$ &
$14.176\pm0.221$ &
$24.252\pm0.174$ \\

1.484 &
$30.512\pm0.760$ &
$12.817\pm0.516$ &
$26.055\pm0.398$ \\

2.330 &
$38.988\pm0.531$ &
$8.632\pm0.101$ &
$31.267\pm0.256$ \\

\hline\hline
\end{tabular}

\end{table}


\section{Results and Discussion}\label{sec4}
\begin{figure}[ht]
    \centering
        \includegraphics[width=1.0\linewidth]{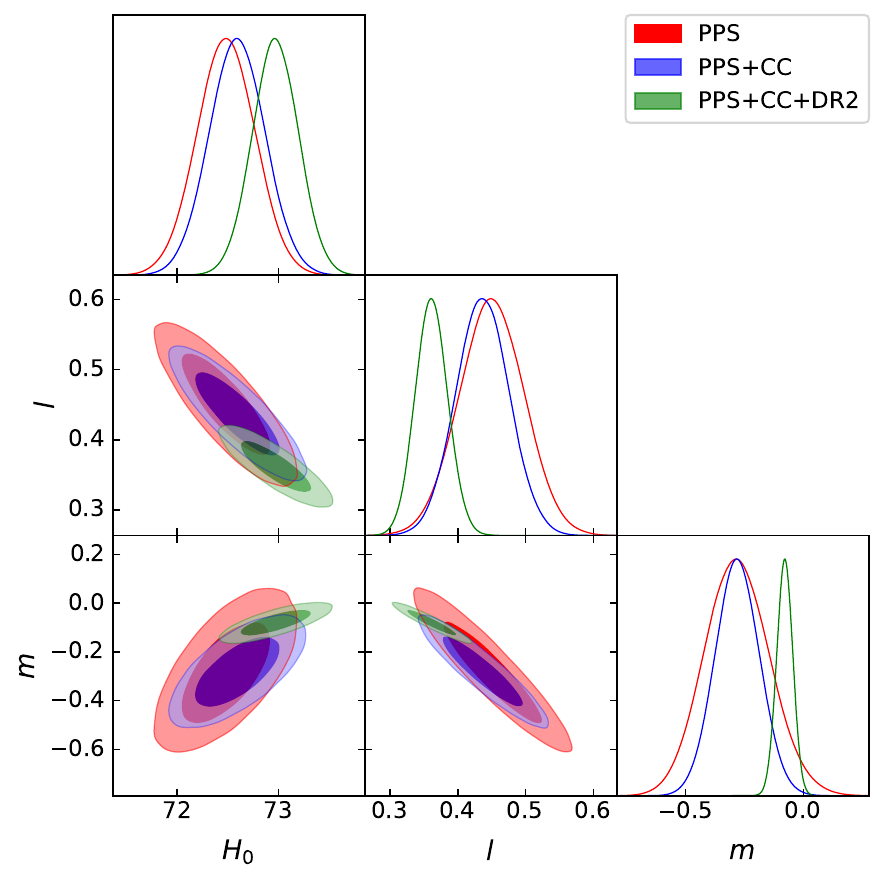}
        \caption{Contour plots showing $1\sigma$ and $2\sigma$ confidence regions of the proposed model with considered datasets.}
        \label{fig:1a}
\end{figure}

\begin{figure}[ht]
    \centering
        \includegraphics[width=1.0\linewidth]{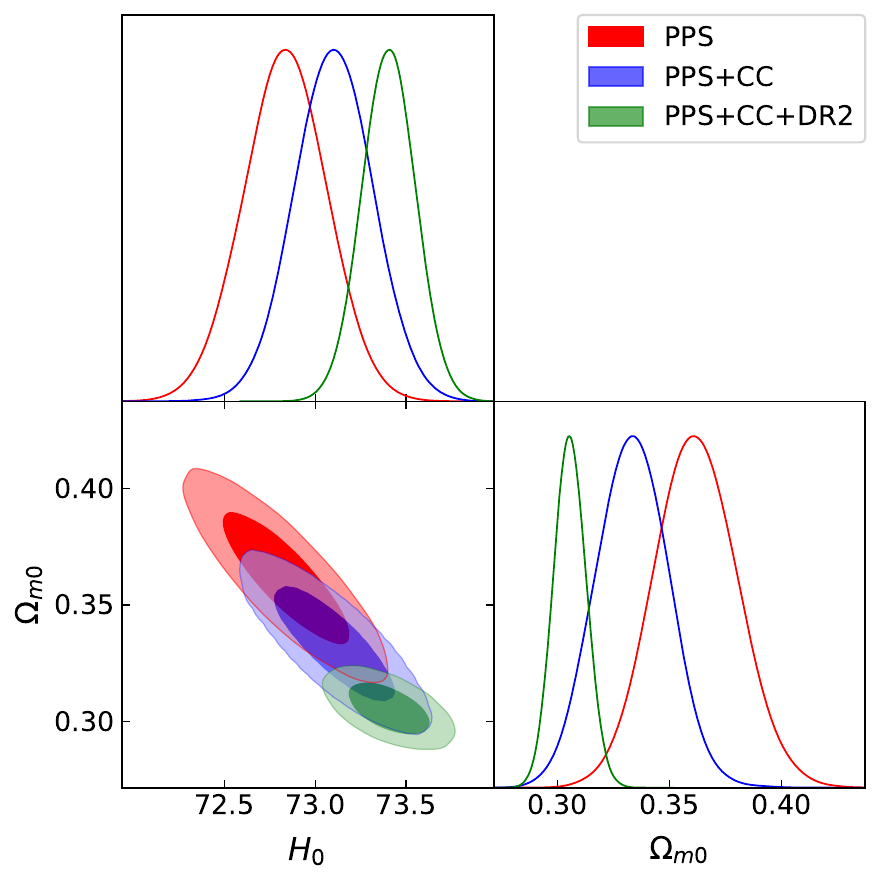}        \caption{Contour plots indicating $1\sigma$ and $2\sigma$ confidence regions for $\Lambda$CDM model with considered datasets.}
        \label{fig:2a}
\end{figure}

In this work, we constrain the free parameter space $(H_0,l,m)$ of the proposed CPL-inspired $Om(z)$ parametrization using the observational datasets PPS, PPS+CC, and PPS+CC+DR2. The best-fit values are obtained through MCMC analysis by maximizing the likelihood function $\mathcal{L}\propto e^{-\chi^2/2}$. Since the datasets are statistically independent, the total likelihood and corresponding chi-square functions are expressed as $\mathcal{L}_{tot}=\mathcal{L}_{CC}\times\mathcal{L}_{PPS}\times\mathcal{L}_{DR2}$ and $\chi^2_{tot}=\chi^2_{CC}+\chi^2_{PPS}+\chi^2_{DR2}$, respectively. The resulting $1\sigma$ and $2\sigma$ confidence contours together with the marginalized distributions are shown in Fig.~\ref{fig:1a} and Fig.~\ref{fig:2a}, while the corresponding best-fit values are summarized in Table~\ref{tab:params}.
\begin{table*}[htbp]
\centering
\caption{Observational constraints on the model parameters $(H_0,l,m)$ obtained from different dataset combinations at $68\%$ and $95\%$ confidence levels.}
\label{tab:params}

\renewcommand{\arraystretch}{1.8}
\setlength{\tabcolsep}{8pt}

\begin{tabular}{l l c c c}
\hline\hline

Dataset &
Model &
$H_0$ {\footnotesize(${\rm km\,s^{-1}\,Mpc^{-1}}$)} &
$l$ &
$m$ \\

\hline

\multirow{2}{*}{PPS}
& Our Model&$72.48\pm0.29^{+0.56}_{-0.56}$
&$0.450\pm0.048^{+0.093}_{-0.094}$
&$-0.28\pm0.14^{+0.28}_{-0.26}$
\\

& $\Lambda$CDM
&$72.84\pm0.23^{+0.45}_{-0.45}$
&$0.362\pm0.019^{+0.037}_{-0.036}$
&$0$
\\

\hline

\multirow{2}{*}{PPS + CC}
& Our Model
&$72.59\pm0.28^{+0.54}_{-0.54}$
&$0.436\pm0.039^{+0.078}_{-0.077}$
&$-0.28\pm0.095^{+0.19}_{-0.18}$
\\

& $\Lambda$CDM
&$73.10\pm0.24^{+0.43}_{-0.43}$
&$0.334\pm0.017^{+0.032}_{-0.031}$
&$0$
\\

\hline

\multirow{2}{*}{PPS + CC + DR2}
& Our Model
&$72.97\pm0.23^{+0.45}_{-0.45}$
&$0.361\pm0.024^{+0.048}_{-0.046}$
&$-0.080\pm0.035^{+0.067}_{-0.068}$
\\

& $\Lambda$CDM
&$73.40\pm0.15^{+0.29}_{-0.29}$
&$0.3056\pm0.0073^{+0.015}_{-0.014}$
&$0$
\\

\hline\hline
\end{tabular}
\end{table*}

The cosmological behaviour of the present model is primarily controlled by the evolution parameter $m$ appearing in the diagnostic relation $Om(z)=l+m\left(\frac{z}{1+z}\right)$. The parameter $l$ determines the present value of the diagnostic and reduces to the matter density parameter in the limiting case $m=0$, corresponding to the flat $\Lambda$CDM scenario. Therefore, any non-zero value of $m$ introduces dynamical evolution in the DE sector.

For the PPS dataset, the obtained Hubble constant is $H_0=72.48\pm0.29~{\rm km\,s^{-1}\,Mpc^{-1}}$, which remains close to the SH0ES local measurement and is slightly lower than the corresponding $\Lambda$CDM value $H_0=72.84\pm0.23~{\rm km\,s^{-1}\,Mpc^{-1}}$. The parameter $l$ is constrained to $l=0.450\pm0.048$, whereas the evolution parameter takes the value $m=-0.28\pm0.14$. The negative value of $m$ indicates a decreasing $Om(z)$ evolution and consequently favors quintessence-like dynamics. However, the relatively large uncertainty still permits mild deviations from the standard cosmological picture.

The inclusion of CC measurements improves the parameter constraints without substantially altering the cosmological evolution. For the PPS+CC dataset, we obtain $H_0=72.59\pm0.28~{\rm km\,s^{-1}\,Mpc^{-1}}$, $l=0.436\pm0.039$, and $m=-0.28\pm0.095$. Compared with PPS alone, the uncertainty in $m$ decreases noticeably while the mean value remains unchanged. This behaviour indicates that the CC dataset primarily acts to tighten the parameter space while preserving the preferred DE evolution.

The tighter constraints are obtained from the combined PPS+CC+DR2 analysis. The inclusion of DR2 measurements significantly reduces the allowed parameter region and shifts the inferred values closer to the $\Lambda$CDM limit. In particular, we obtain $H_0=72.97\pm0.23~{\rm km\,s^{-1}\,Mpc^{-1}}$, $l=0.361\pm0.024$, and $m=-0.080\pm0.035$. The substantial reduction in the magnitude of $m$ indicates that large deviations from constant $Om(z)$ evolution are strongly disfavored once BAO information is incorporated. Although the evolution parameter remains negative, its proximity to zero suggests that the model remains observationally close to the $\Lambda$CDM framework.

The contour plots shown in Fig.~\ref{fig:1a} reveal the correlations among the parameters $(H_0,l,m)$ for different dataset combinations. A clear anti-correlation between $l$ and $m$ is observed throughout the analysis, indicating that larger values of the present diagnostic parameter $l$ correspond to smaller evolution in the $Om(z)$ profile, as quantified by $m$. In contrast, the parameter $H_0$ exhibits a positive correlation with $m$, suggesting that higher expansion rates are associated with larger values of the present $Om(z)$ amplitude.

For comparison, the corresponding $\Lambda$CDM contour plots shown in Fig.~\ref{fig:2a} illustrate the constraints on $(H_0,\Omega_{m0})$. Similar to the present model, the inclusion of additional datasets progressively reduces the allowed parameter space and tightens the marginalized distributions. The matter density parameter systematically decreases from $\Omega_{m0}=0.362$ for PPS to $\Omega_{m0}=0.3056$ for the combined PPS+CC+DR2 analysis, accompanied by tighter parameter constraints. This behaviour indicates the improved constraining power of the joint observational analysis and reflects the role of DR2 data in reducing parameter degeneracies.

Another important feature observed from the contour plots is the progressive reduction of parameter degeneracies with the inclusion of additional datasets. The parameter $l$ systematically decreases from $l\approx0.450$ for PPS to $l\approx0.361$ for the combined PPS+CC+DR2 analysis. Similarly, the uncertainty in $m$ decreases from approximately $0.14$ to $0.035$, confirming the constraining power of the DR2 dataset.

From the physical point of view, the negative values of $m$ obtained for all dataset combinations imply a decreasing evolution of the $Om(z)$ diagnostic with redshift. This behaviour is generally associated with quintessence-like DE dynamics $(\omega>-1)$. Nevertheless, the joint analysis progressively shifts $m$ toward the $\Lambda$CDM limit ($m=0$), indicating that the preferred evolution remains mild.

The Hubble constant obtained in the present analysis lies within the interval $72.48\lesssim H_0\lesssim72.97~{\rm km\,s^{-1}\,Mpc^{-1}}$, depending on the dataset combination. These values remain consistent with the SH0ES determination $H_0=73.04\pm1.04~{\rm km\,s^{-1}\,Mpc^{-1}}$~\cite{SupernovaSearchTeam:2004lze} and are comparatively higher than the Planck 2018 estimate $H_0=67.4\pm0.5~{\rm km\,s^{-1}\,Mpc^{-1}}$~\cite{aghanim2020planck}.

Overall, the proposed parametrization provides a statistically consistent description of the late-time expansion history of the Universe. The combined PPS+CC+DR2 dataset yields tighter constraints and favors a cosmological scenario characterized by lower matter density and weak evolution in the $Om(z)$ diagnostic. Although the model allows dynamical departures through non-zero values of $m$, the observational constraints indicate that the preferred evolution remains close to the $\Lambda$CDM limit, supporting the viability of the proposed $Om(z)$ cosmology.

\subsection{Statistical Analysis}

To assess the goodness of fit of the proposed model with observational datasets, we first analyze the minimum chi-square $(\chi^2_{min})$ and reduced chi-square $(\chi^2_r)$. The corresponding values for our Model and the $\Lambda$CDM scenario are summarized in Table~\ref{tab:stats}.
\begin{table}[ht]
\centering
\caption{Observational constraints on statistical parameters for our Model and $\Lambda$CDM from different dataset combinations.}
\label{tab:stats}
\renewcommand{\arraystretch}{1.5}

\resizebox{\columnwidth}{!}{%
\begin{tabular}{l c c c c c}
\hline\hline
Dataset & Model & $\chi^2_{min}$ & $\chi^2_r$ & AIC & BIC \\
\hline

\multirow{2}{*}{PPS}
& Our Model & 1748.33 & 1.03 & 1754.33 & 1770.65 \\
& $\Lambda$CDM & 1752.48 & 1.03 & 1756.48 & 1767.35 \\

\hline

\multirow{2}{*}{PPS + CC}
& Our Model & 1773.67 & 1.03 & 1779.67 & 1795.99 \\
& $\Lambda$CDM & 1782.31 & 1.03 & 1786.31 & 1797.18 \\

\hline

\multirow{2}{*}{PPS + CC + DR2}
& Our Model & 1797.19 & 1.03 & 1805.19 & 1827.04 \\
& $\Lambda$CDM & 1801.00 & 1.04 & 1807.00 & 1823.00 \\

\hline\hline
\end{tabular}}
\end{table}

The minimum chi-square is obtained from the maximum likelihood estimation as $\chi^2_{min}=-2\ln(\mathcal{L}_{max})$, where $\mathcal{L}_{max}$ denotes the maximum likelihood function. The reduced chi-square is defined as $\chi^2_r=\chi^2_{min}/(N-k)$, where $N$ and $k$ represent the total number of observational data points and free model parameters, respectively. In general, values of $\chi^2_r\approx1$ indicate good agreement between the model and observational data.

For our Model, the minimum chi-square values are obtained as $\chi^2_{min}=1748.33$, $1773.67$, and $1797.19$ for the PPS, PPS+CC, and PPS+CC+DR2 datasets, respectively. The corresponding reduced chi-square values remain nearly constant with $\chi^2_r\approx1.03$ for all dataset combinations, indicating good agreement with observations. For comparison, the $\Lambda$CDM model yields slightly larger values of $\chi^2_{min}$, namely $1752.48$, $1782.31$, and $1801.00$, suggesting a marginally improved fit of the proposed model.

For model comparison, we further employ the Akaike Information Criterion (AIC)~\cite{akaike2003new,
burnham2011aic} and Bayesian Information Criterion (BIC)~\cite{schwarz1978estimating} defined as $AIC=\chi^2_{min}+2k$ and $BIC=\chi^2_{min}+k\ln N$, respectively. Smaller values of AIC and BIC generally indicate a statistically preferred model. The corresponding values obtained for our Model and the $\Lambda$CDM scenario are listed in Table~\ref{tab:stats}.

The proposed model yields AIC values of $1754.33$, $1779.67$, and $1805.19$ for PPS, PPS+CC, and PPS+CC+DR2 datasets, respectively, which are lower than the corresponding $\Lambda$CDM values of $1756.48$, $1786.31$, and $1807.00$. Similarly, the BIC values remain comparatively lower for our Model in most dataset combinations. Therefore, the statistical analysis indicates that the proposed $Om(z)$ parametrization provides a satisfactory fit to the observational data and remains statistically competitive with the $\Lambda$CDM framework.

Thus, the statistical analysis indicates that the proposed $Om(z)$ parametrization provides a satisfactory fit to the observational data and remains statistically competitive with the standard $\Lambda$CDM framework. The lower $\chi^2_{min}$ and AIC values obtained for our Model suggest a mild statistical preference over $\Lambda$CDM. Since the model reduces to the standard cosmological scenario in the limit $m=0$, while allowing dynamical evolution through non-zero values of $m$, the present parametrization may be regarded as a viable extension of the $\Lambda$CDM cosmology.


\subsection{Behaviour of Deceleration Parameter}
The expansion dynamics of the Universe can be characterized through the deceleration parameter, which distinguishes accelerated and decelerated expansion phases. It is defined as
\begin{equation}
q(z)=-1+\frac{1+z}{H(z)}\frac{dH(z)}{dz},
\end{equation}
where $H(z)$ is the Hubble parameter and the derivative is taken with respect to redshift. Positive values of $q(z)$ correspond to a decelerating Universe, whereas negative values indicate accelerated expansion.

\begin{figure}[ht]
    \centering
        \includegraphics[width=1\linewidth]{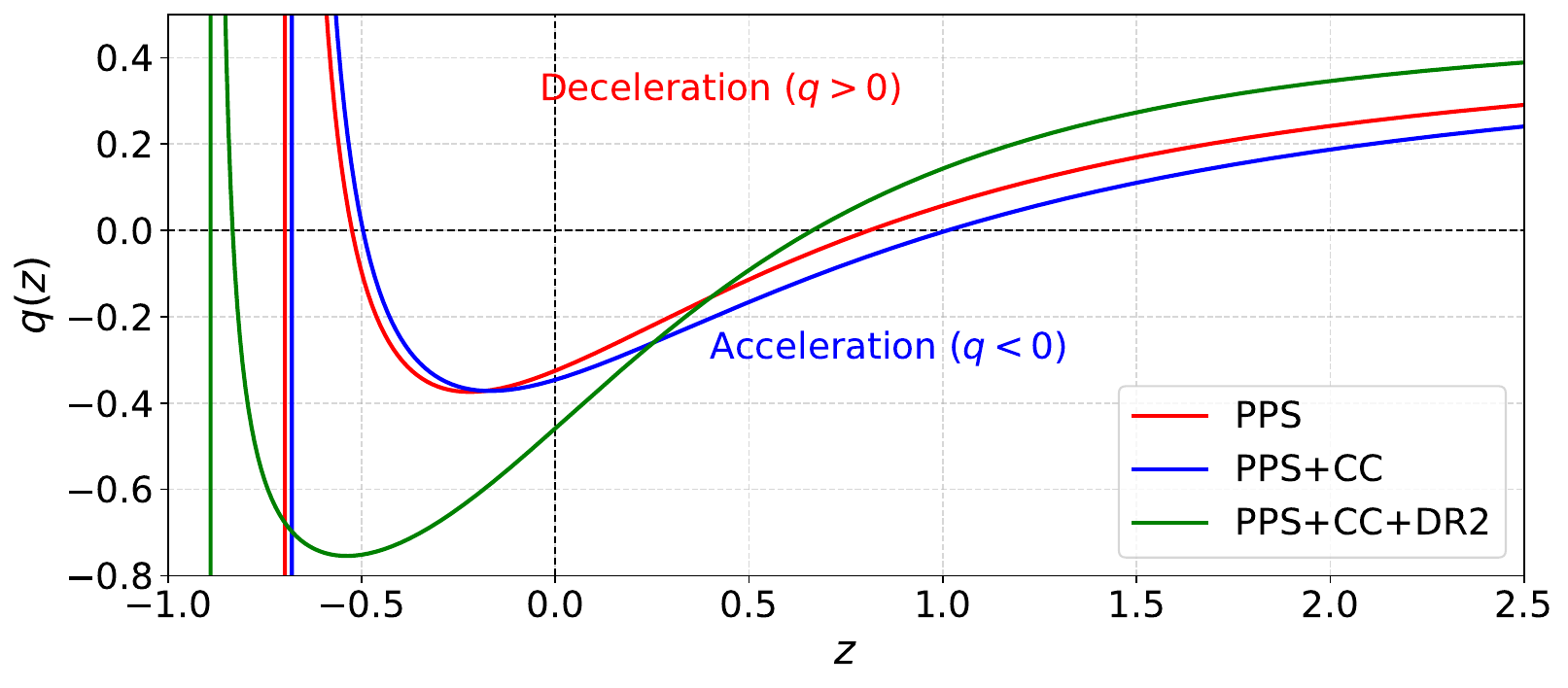}
        \caption{Evolution of deceleration parameter $q(z)$ for considered model.}
    \label{fig:q}
\end{figure}

Using the Hubble parameter obtained from Eq.~(\ref{Hz_model}), the deceleration parameter for the proposed $Om(z)$ parametrization becomes
\begin{equation}
q(z)=-1+ \frac{m\left(\dfrac{(1+z)^3-1}{1+z}\right) +3(1+z)^2\left[A(z)\right]}{2\left[1+\left(l+m\left(\frac{z}{1+z}\right)\right) \left((1+z)^3-1\right)\right]},
\end{equation}
where $A(z)=l(1+z) +3mz$.
The evolution of $q(z)$ as a function of redshift is shown in Fig.~\ref{fig:q} for different dataset combinations. The obtained evolution exhibits multiple crossings between accelerated and decelerated phases, indicating a non-trivial cosmological behaviour.

The present values of the deceleration parameter $(q_0)$ and the corresponding transition redshifts are summarized in Table~\ref{tab:physical}. For the PPS dataset, the present value is obtained as $q_0=-0.325$. With the inclusion of CC data, the value shifts slightly to $q_0=-0.346$, while the strongest constraint is obtained from the combined PPS+CC+DR2 analysis with $q_0=-0.459$. The negative values obtained for all dataset combinations confirm that the Universe is currently undergoing accelerated expansion.

\begin{table}[ht]
\centering
\caption{Observational constraints on physical parameters for our Model from different dataset combinations.}
\label{tab:physical}
\renewcommand{\arraystretch}{1.5}

\resizebox{\columnwidth}{!}{%
\begin{tabular}{l c c c c c}
\hline\hline
Dataset & $q_0$ & $z_{tr}$ & $z_{tr1}$ & $z_{tr2}$ & $\omega_0$ \\
\hline

PPS & $-0.325$ & $0.810$ & $-0.525$ & $-0.698$ & $-0.550$ \\

PPS + CC & $-0.346$ & $1.012$ & $-0.496$ & $-0.680$ & $-0.564$ \\

PPS + CC + DR2 & $-0.459$ & $0.664$ & $-0.833$ & $-0.890$ & $-0.639$ \\

\hline\hline
\end{tabular}}
\end{table}

An important feature of the present model is the existence of three transition redshifts. The first transition, denoted by $z_{tr}$, corresponds to the standard transition from a decelerated phase to accelerated expansion at late times. For the PPS dataset, this transition occurs at $z_{tr}=0.810$, indicating the onset of acceleration around this epoch. The inclusion of CC data shifts the transition to $z_{tr}=1.012$, suggesting an earlier onset of cosmic acceleration. However, the combined PPS+CC+DR2 dataset yields $z_{tr}=0.664$, implying a comparatively later transition to the accelerated phase.

The second transition, represented by $z_{tr1}$, corresponds to a future evolution from acceleration to deceleration. The transition occurs at $z_{tr1}=-0.698$ for PPS and shifts slightly to $z_{tr1}=-0.680$ for PPS+CC. The combined PPS+CC+DR2 analysis further moves this transition to $z_{tr1}=-0.890$, indicating that the accelerated phase persists for a longer duration before entering a future decelerated epoch.

Furthermore, the model predicts another future transition denoted by $z_{tr2}$, where the cosmic expansion again changes from deceleration to acceleration. This transition occurs at $z_{tr2}=-0.525$ for PPS and $z_{tr2}=-0.496$ for PPS+CC, while the inclusion of DR2 shifts it to $z_{tr2}=-0.833$. Since these transition points also lie in the future region $(z<0)$, the model predicts repeated switching between accelerated and decelerated phases during future cosmic evolution.

Therefore, unlike the standard $\Lambda$CDM scenario, which predicts a monotonic late-time acceleration, the present $Om(z)$ parametrization allows multiple transition phases in the future evolution. Nevertheless, the current cosmological epoch remains consistently accelerated for all observational datasets.
Therefore, the evolution of $q(z)$ indicates a smooth transition from the matter-dominated decelerating phase to the present accelerated epoch, while allowing additional future transitions governed by the dynamical behaviour of the proposed $Om(z)$ parametrization.

\subsection{Total EoS and Energy Density}
\begin{figure}[ht]
        \centering
        \includegraphics[width=\linewidth]{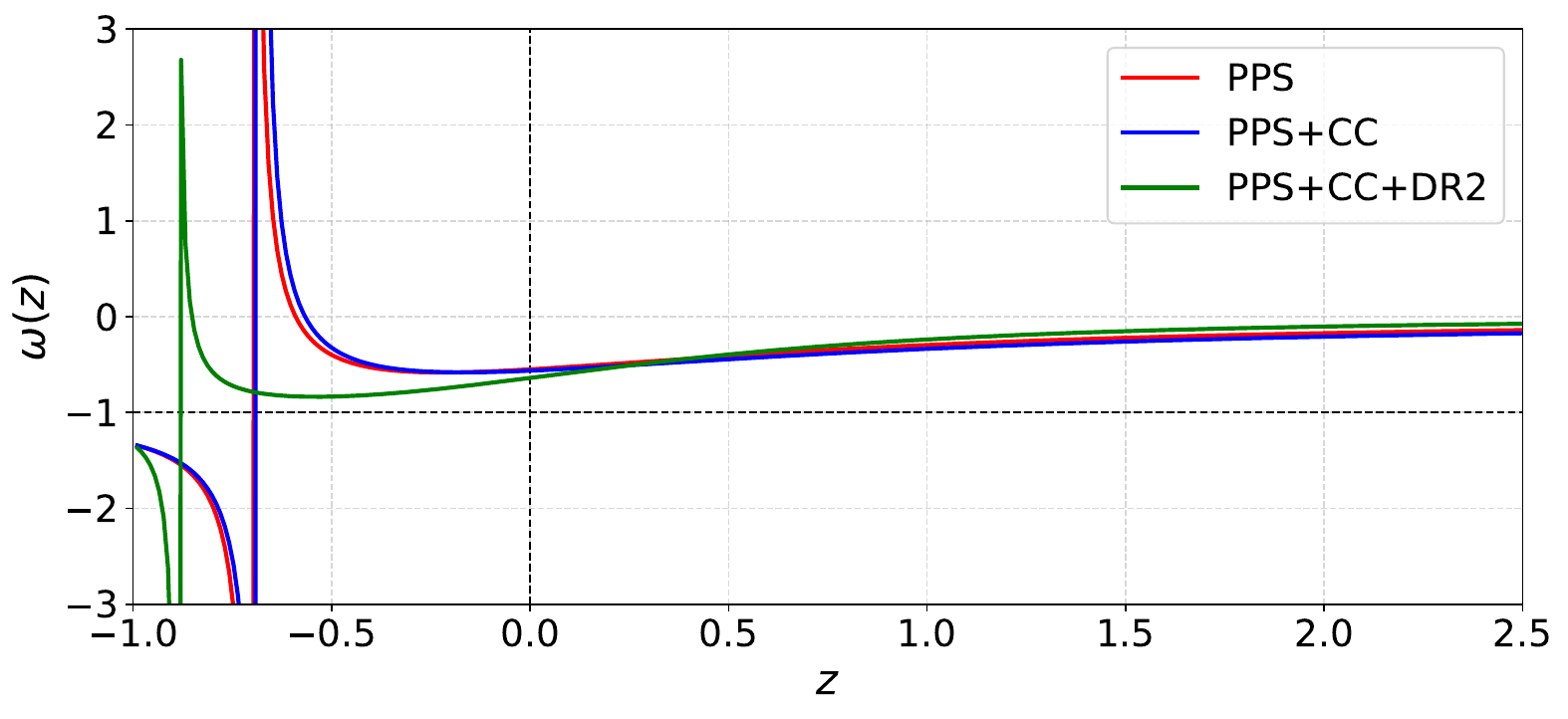}
        \caption{Evolution of total EoS parameter $\omega_{tot}$ for considered model.}
        \label{fig:w}
    \end{figure}

To further investigate the physical behaviour of the proposed cosmological model, we analyze the evolution of the total equation of state (EoS) parameter and the corresponding energy density. The total EoS parameter is defined as
\begin{equation}
\omega(z)=\frac{p}{\rho}=-\frac{2\dot{H}+3H^2}{3H^2}=-1-\frac{2}{3}\frac{\dot H}{H^2},
\end{equation}
which relates the cosmic expansion directly to the Hubble dynamics. Using the relation between the deceleration parameter and the EoS parameter, the above expression can be written as
\begin{equation}
\omega(z)=\frac{2q(z)-1}{3}.
\end{equation}
Substituting the obtained expression of the deceleration parameter into the above relation, we obtain
\begin{equation}
\omega(z)=-1+\frac{1}{3}\frac{m\left(\dfrac{(1+z)^3-1}{1+z}\right)+3(1+z)^2\left[A(z)\right]}{1+\left(l+m\left(\frac{z}{1+z}\right)\right)\left((1+z)^3-1\right)}.
\end{equation}
The evolution of the total EoS parameter is shown in Fig.~\ref{fig:w} for different dataset combinations. The obtained behaviour indicates that the EoS evolves from values close to the matter-dominated regime at higher redshifts toward negative values at lower redshifts, confirming the transition to an accelerated expansion phase.

The present values of the EoS parameter are summarized in Table~\ref{tab:physical}. For the PPS dataset, the present value is obtained as $\omega_0=-0.550$, while the inclusion of CC data slightly shifts it to $\omega_0=-0.564$. The combined PPS+CC+DR2 analysis yields $\omega_0=-0.639$, corresponding to the strongest deviation toward negative values.
For all dataset combinations, the present EoS remains within the interval $-1<\omega_0<-\frac{1}{3}$, indicating quintessence-like behaviour and confirming the accelerated expansion of the present Universe. Furthermore, the obtained evolution of DE EoS at present does not cross the phantom boundary $\omega=-1$, ensuring a physically viable cosmological evolution.

At higher redshifts, the EoS parameter evolves toward less negative values, indicating the matter-dominated phase of the early Universe. As the Universe evolves toward lower redshifts, $\omega(z)$ gradually decreases and reaches the present accelerated epoch, remaining within the quintessence region $(-1<\omega<-1/3)$.
However, the future evolution $(z<0)$ exhibits a different behaviour. The EoS parameter crosses the phantom divide line $(\omega=-1)$ and evolves into the phantom regime $(\omega<-1)$ before exhibiting a divergence near the transition epoch. Therefore, the model predicts a transition from matter domination to the present quintessence-like accelerated phase, followed by phantom evolution in the future Universe.
\begin{figure}[ht]
    \centering
        \includegraphics[width=1\linewidth]{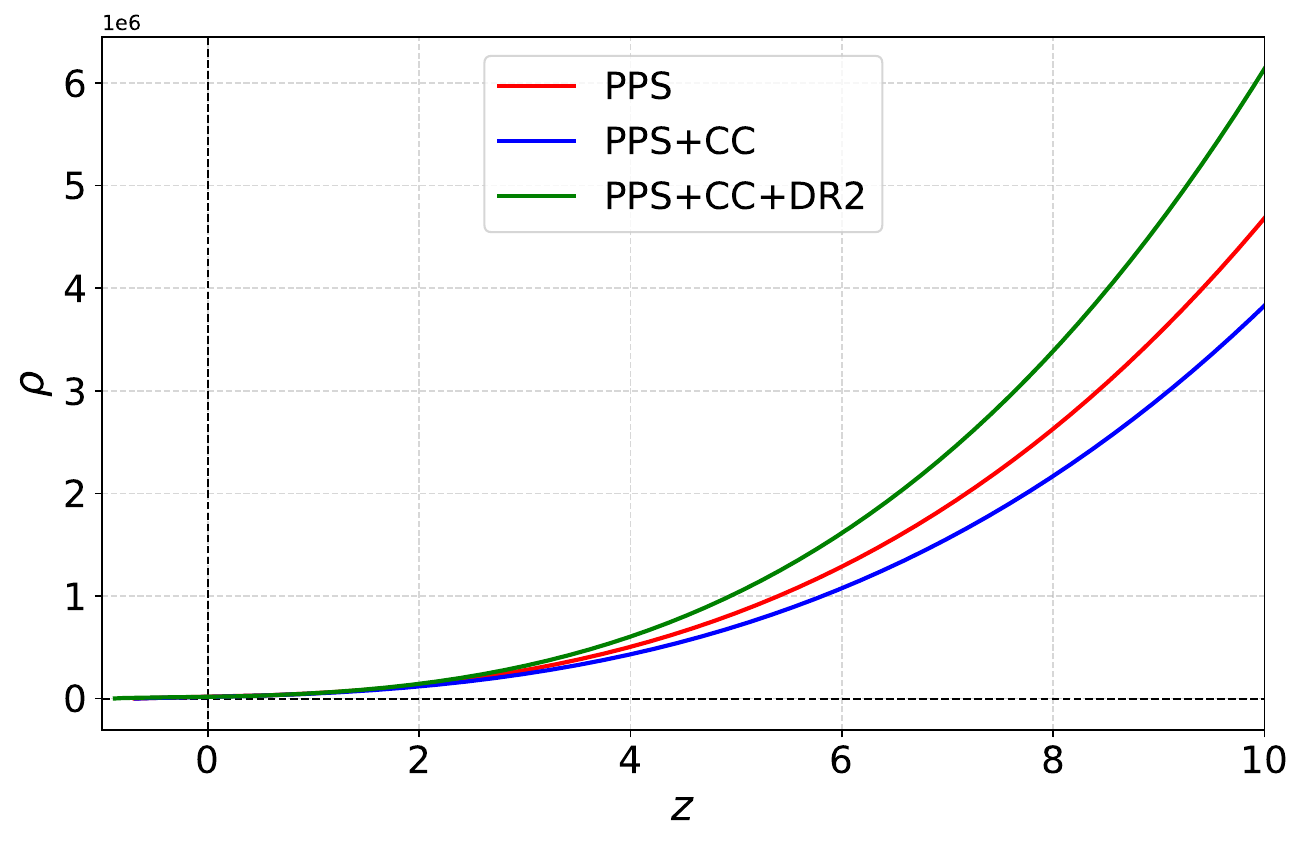}
        \caption{Evolution of total energy density $\rho_{tot}$ for considered model.}
    \label{fig:rho}
\end{figure}

The corresponding total energy density is obtained from the Friedmann equation as $\rho(z)=3H^2(z)$. The evolution of the total energy density is shown in Fig.~\ref{fig:rho}. The energy density increases toward higher redshift due to the increasing matter contribution at earlier epochs and decreases smoothly toward lower redshift because of cosmic expansion. This behaviour remains consistent for all dataset combinations and indicates a stable cosmological evolution.

Overall, the evolution of $\omega(z)$ and $\rho(z)$ supports a cosmological picture in which the Universe evolves from a matter-dominated decelerating phase in the past to the presently accelerated epoch, while maintaining quintessence-like behaviour throughout the cosmic evolution.

\subsection{$Om(z)$ Diagnostic}
\begin{figure}[ht]
    \centering
        \includegraphics[width=1\linewidth]{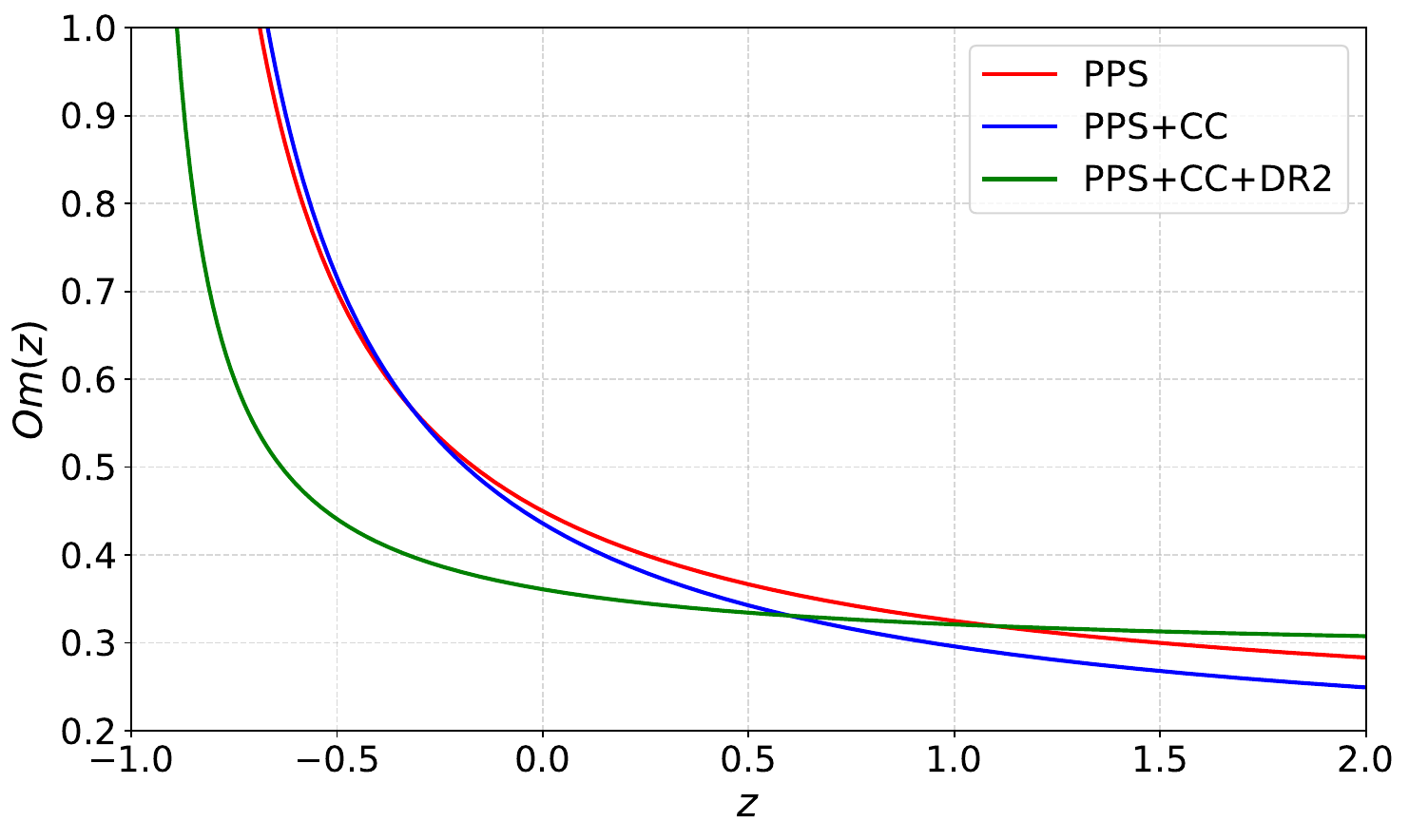}
        \caption{Evolution of $Om(z)$ for considered model.}
    \label{fig:om}
\end{figure}

The evolution of the proposed $Om(z)$ diagnostic for different dataset combinations is shown in Fig.~\ref{fig:om}. The obtained profiles exhibit a decreasing behaviour with redshift throughout the evolution, indicating quintessence-like dynamics.
The PPS dataset exhibits a relatively stronger evolution of the diagnostic. With the inclusion of CC data, the decreasing behaviour is preserved while slightly modifying the evolution profile. The combined PPS+CC+DR2 analysis exhibits a comparatively weaker evolution than the PPS and PPS+CC datasets, although the decreasing trend remains unchanged.

The present values of the diagnostic progressively decrease from PPS to the combined analysis, reflecting the influence of additional observational datasets on the cosmological evolution. Furthermore, the values at the future epoch $(z=-1)$ also distinguish the dataset combinations and provide information regarding their deviations in the cosmic evolution.
Since the proposed parametrization reduces to a constant diagnostic in the limit $m=0$, corresponding to the flat $\Lambda$CDM scenario, the obtained evolution indicates that the present model remains dynamically evolving for all dataset combinations.

\section{Final Remarks}\label{sec5}

In this work, we proposed and investigated a new CPL-inspired parametrization of the $Om(z)$ diagnostic given by $Om(z)=l+m\left(\dfrac{z}{1+z}\right)$, which introduces a dynamical evolution through the free parameter $m$ while reducing to the flat $\Lambda$CDM cosmology in the limit $m=0$. The observational viability of the model was examined using recent late-time datasets including PPS, CC, and DESI BAO DR2 observations. Furthermore, the cosmological implications of the obtained evolution were analyzed through the deceleration parameter, total EoS parameter, total energy density, and the $Om(z)$ diagnostic. The main findings of the present work are summarized as follows:

\begin{itemize}

\item[(i)] The Hubble constant is constrained within the interval $72.48 \lesssim H_0 \lesssim 72.97~\mathrm{km\,s^{-1}\,Mpc^{-1}}$, remaining consistent with local SH0ES measurements. The parameter $l$ decreases systematically with the inclusion of additional datasets, while the evolution parameter remains negative for all dataset combinations.

\item[(ii)]The inclusion of DR2 data significantly tightens the parameter space and shifts the evolution closer to the $\Lambda$CDM limit. The evolution parameter shifts toward smaller values, indicating a reduced dynamical evolution in comparison with the PPS and PPS+CC analysis.

\item[(iii)] The statistical analysis indicates good agreement between the model and observational data with $\chi_r^2\approx1.03$ for all dataset combinations. Furthermore, the model yields slightly lower $\chi^2_{\min}$ and AIC values compared with the $\Lambda$CDM scenario, suggesting a mild statistical preference. Since the model reduces to the standard cosmology in the limit $m=0$, it may be regarded as a viable extension of the $\Lambda$CDM framework.

\item[(iv)] The physical analysis confirms that the Universe remains in the present accelerated phase with negative values of the deceleration parameter for all datasets. The evolution of $q(z)$ indicates a transition from the past matter-dominated decelerated epoch to the present accelerated phase, while the model further predicts additional future transitions between acceleration and deceleration.

\item[(v)] The total EoS parameter remains within the quintessence region at the present epoch with $-1<\omega_0<-\dfrac{1}{3}$, confirming the accelerated expansion of the Universe. However, the future evolution indicates a crossing of the phantom divide line followed by phantom behaviour before divergence near the transition epoch.

\item[(vi)] The evolution of the proposed $Om(z)$ diagnostic exhibits a decreasing behaviour for all dataset combinations, indicating quintessence-like dynamics. Although the combined PPS+CC+DR2 analysis shifts the evolution closer to the constant $\Lambda$CDM behaviour, the model remains dynamically evolving throughout the cosmic evolution.

\end{itemize}

\noindent In summary, the proposed CPL-inspired $Om(z)$ parametrization provides a statistically consistent and observationally viable description of the late-time Universe. The obtained results indicate mild deviations from the standard $\Lambda$CDM cosmology while remaining compatible with current observations. Since the present parametrization exhibits dynamical evolution and non-trivial future behaviour, it may also be extended and investigated within the framework of modified theories of gravity. Furthermore, future high-precision cosmological observations~\cite{DESI:2025fxa, Spec-S5:2025uom, Euclid:2024yrr, LSST:2008ijt, CMB-S4:2016ple} may provide tighter constraints on the evolution parameter and offer further insights into the dynamical nature of DE.

\bigskip
\noindent\textbf{Data availability statement} We employed publicly available CC, Pantheon+SH0ES, and DESI BAO DR2 data presented in this study. The Pantheon+SH0ES data compilation (distance moduli and covariance matrices), which is publicly available on GitHub: \url{https://github.com/PantheonPlusSH0ES/DataRelease}. The sources of the CC and DESI BAO DR2 datasets are provided in the Data section. No additional data were used in this study.

\bigskip
\noindent\textbf{Declaration of competing interest} The authors declare no competing interests.\\

\bibliographystyle{utphys}
\bibliography{references}

\end{document}